\documentclass[sigconf]{acmart}

\copyrightyear{2022}
\acmYear{2022}
\setcopyright{rightsretained}\acmConference[CHAI '22]{Proceedings of the HAI'22 workshop on
Cognitive Human-agent Interaction}{December 5, 2022}{Christchurch, New Zealand}
\acmBooktitle{Proceedings of the HAI'22 workshop on
Cognitive Human-agent Interaction}
\acmPrice{}
\acmDOI{}
\acmISBN{}

\begin{document}

\title{Emotion in Cognitive Architecture:\\ Emergent Properties from Interactions with Human Emotion}

\author{Junya Morita}
\email{j-morita@inf.shizuoka.ac.jp}
\orcid{1234-5678-9012}
\affiliation{%
  \institution{Shizuoka University}
  \streetaddress{3-5-1 Johoku, Naka-kuk}
  \city{Hamamatsu}
  \state{Shizuoka}
  \country{Japan}
  \postcode{432-8015}
}

\renewcommand{\shortauthors}{Morita}

\begin{abstract}
This document presents endeavors to represent emotion in a computational cognitive architecture. The first part introduces research organizing with two axes of emotional affect: pleasantness and arousal. Following this basic of emotional components, the document discusses an aspect of emergent properties of emotion, showing interaction studies with human users. With these past author's studies, the document concludes that the advantage of the cognitive human-agent interaction approach is in representing human internal states and processes. 

\end{abstract}

\ccsdesc[300]{Human-centered computing~HCI theory, concepts and models}



\keywords{cognitive architecture, emotion, arousal, valence}

\maketitle

\section{Introduction}
This document begins by asking the following question:
\begin{quote}
{\it How can emotion emerge in a computational system?}
\end{quote}
This is a kind of ultimate question that has attracted enormous numbers of scientists and engineers, including the author himself. 
Toward the complete answer to the question, the author has developed several models of mental functions (possible components of emotion process) in ACT-R (Adaptive Control of Thought-Rational \cite{Anderson:2007}), which is one of the most widely used cognitive architectures in the world. 

Cognitive architectures generally integrate knowledge concerning the human mind in the form of computer programs. The knowledge accumulated in ACT-R ranges from perceptual-motor components to abstract and goal-related concepts. Varieties of mental functions are controlled by symbols stored in modules (corresponding brain regions) and subsymbolic parameters (corresponding neurotransmitters). By utilizing these, this architecture aims to realize human-level activities in every field of human life. 

The author considers that the above characteristic of the architecture is crucial to answering the question. Thus, this document presents the author's works utilizing ACT-R as ingredients of discussions on the conditions for enabling emotion in a computational system.

\section{Issues of Emotional Process}

As the background of the discussion, this section presents the author's view on the emotion process. It is assumed that the emotion process relies on subcortical brain regions, such as the amygdala, insula, and basal ganglia. The process was considered to be the product of an adaptation to ancestors' surrounding environments, because these regions were formed early in the evolution of the brain \cite{damasio,ledoux2020deep}. However, modern human emotion is more complex than the purely physical processes in the following senses:

\begin{enumerate}
\item Emotion is not a static entity but rather emergent properties accompanied by dynamic interaction with the environment. This means a human's emotional response always fluctuates.
\item Our environment has drastically changed from the environment in which our ancestors lived. Therefore, many emotional mechanisms have become maladaptive in the modern age.
\item  This biological process is somehow modulated by an intentional strategy, as many theories of emotion  have suggested \cite{damasio,barrett2017emotions}. Therefore, there is room for intervention in the emotion process by our will or technology.
\end{enumerate}

\section{Representing Emotion in ACT-R}

Even through such complexities, it is possible to model the basis of emotion (affect) using the fundamental axes, arousal and pleasantness, as presented in Russel \cite{russell2003core}'s circumplex model (Fig 1). The present author's approach begins with these simple components. The following studies represent these components with primitive cognitive functions, such as activation noise and pattern matching,  implemented in ACT-R.

\begin{figure}[t]
\centering
\includegraphics[bb=0 0 445 406, width=0.8\columnwidth]{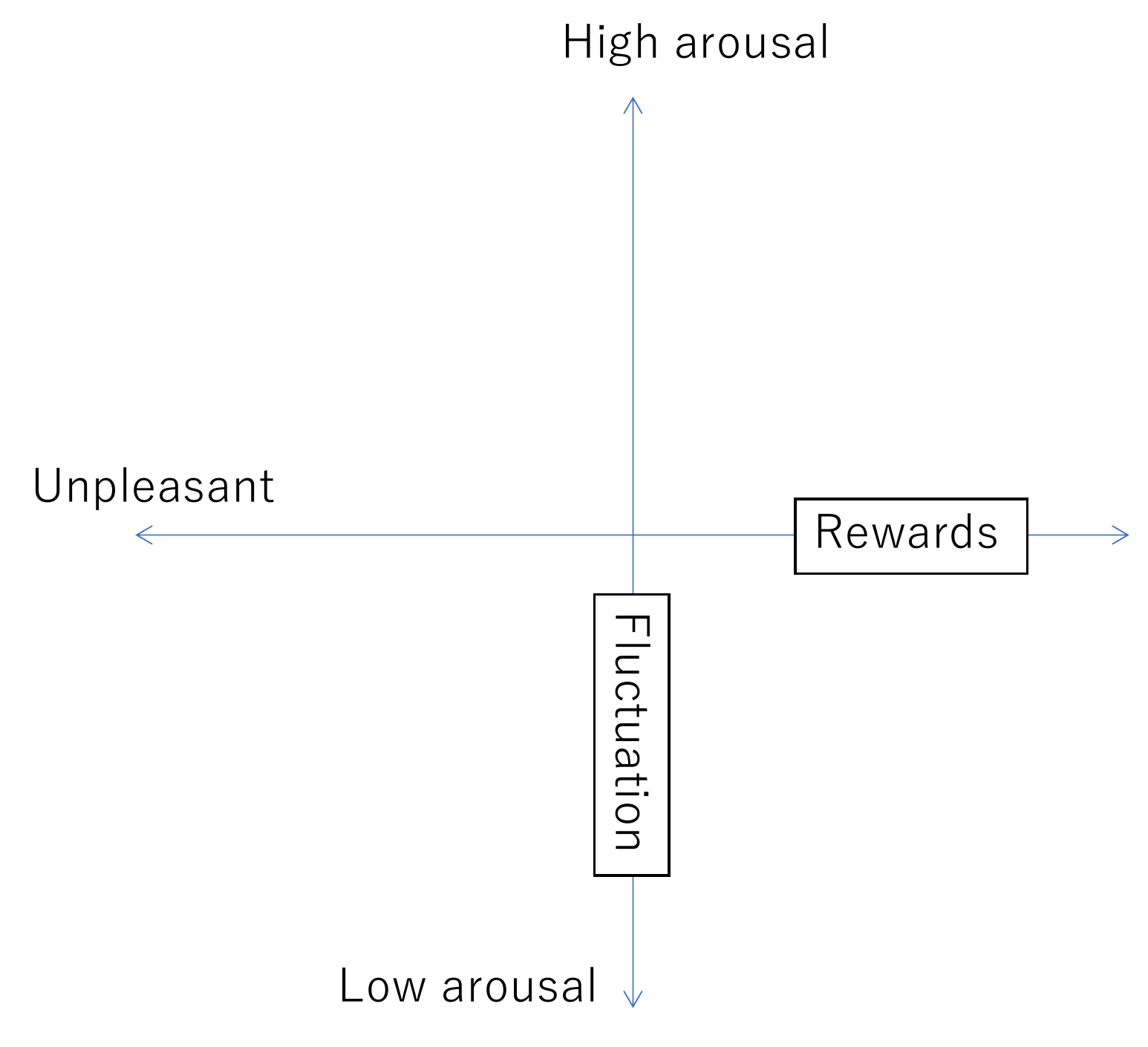}
\caption{Axes of emotional affect and model parameters.}
\label{fig:hypothesis}
\end{figure}

\subsection{Representation of Emotional Arousal}
According to a dictionary in psychology \cite{vandenbos2007apa}, arousal is defined as follows: 
\begin{quote}
 a state of physiological activation or cortical responsiveness, associated with sensory stimulation and activation of fibers from the reticular activating system.
\end{quote}
As indicated by this definition, past researchers have frequently connected arousal with the activation (attentional) process, especially the degree of concentration (e.g., \cite{landers1980arousal}). In a concentrated situation, humans can continue monotonous tasks accurately. However, as such a process long lasts long, people get bored and begin to think about things outside of the task (i.e., mind-wandering). 

From this phenomenon and the consideration that emotional arousal relates to biological fluctuations affecting the cognitive process, the author and their colleagues have represented arousal as the noise factor for memory activation in ACT-R. Controlling this subsymbolic parameter, the authors have demonstrated changes in memory recollection \cite{morita2016modelbased} and task goal switching \cite{nagashima}. This view is consistent with the discussion in which emotion is a modulator of cognitive architecture \cite{ritter2009two,dancy2015using}.

\subsection{Representation of Pleasantness}
People feel joy when receiving a reward. Thus, the pleasantness axis can be considered as a reward in reinforcement learning. Various triggering events for rewards can be assumed in the real world. Among them, internally generated ones are important to developing autonomous agents. 

Based on the above assumptions, Nagashima et al. \cite{iccm2021nagashima,pre1b} developed a model of intrinsic motivation, assuming a pattern discovery to be a source of curiosity. In the ACT-R model, patterns in the data are discovered with the process of pattern matching, and the experience of the pattern matching is utilized to build proceduralized rules (the compilation of production rules). As a task proceeds, opportunities for pattern matching (internal rewards) gradually decrease. Thus, the model can explain how intrinsic motivation decreases with experience and increases with discovering novel patterns. This pattern-discovery-focused view of intrinsic motivation is consistent with discussions in the entertainment industry \cite{koster2013theory}. In addition, it is supported by the theory emphasizing the role of pattern-seeking in the history of human civilization \cite{baron2020pattern}.

\section{Needs of Interaction}
From the discussion so far, the importance of environment for emotion stands out. 
Capturing the full complexities of the real-world dynamics of the two mechanisms (arousal and pleasantness) requires environmental changes. Among various environmental factors, the existence of other organisms seems most crucial. In other words, the author considers that human emotion can be modeled only when the computational system actually interacts with humans. By interacting with humans, the system is able to learn human's emotion generation and expression. 

Based on this consideration, the author and colleagues have developed several interactive systems \cite{itabashi2020,morita2022regulating}, implementing an emotion model as a component.
Those systems receive the user's biological signals such as heart rates to automatically modulate the abovementioned ACT-R parameters (noises and rewards) for guiding the user's emotion to an optimal state. Especially, Morita et al. \cite{morita2022regulating} demonstrated that a web advertisement system containing an ACT-R memory model could prevent human repetitive thinking (rumination) when the model behaves in a counterbalanced manner (Fig 2). The author believes that such a system will eventually lead to a new human homeostatic process with the help of artificial emotional systems.

\begin{figure}[t]
\centering
\includegraphics[bb=0 0 363 327,width=0.8\columnwidth]{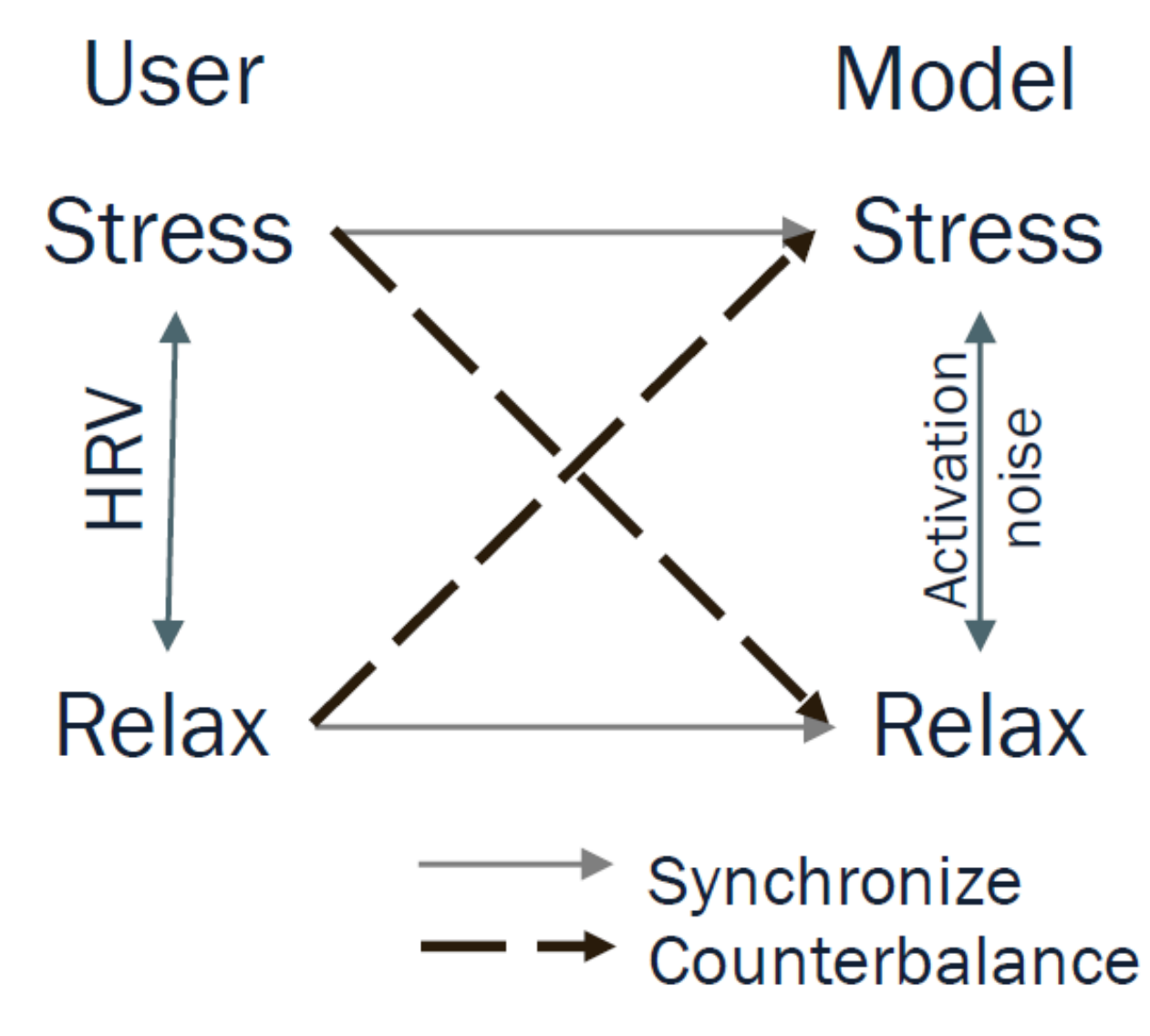}
\caption{Framework of interacting human emotion with machine emotion \cite{morita2022regulating}}
\label{fig:hypothesis}
\end{figure}

\section{Conclusion}
This document presents the author's attempts to answer the ultimate question about the computational conditions that enable emotion. Future work must represent the ideas presented in the document as a general framework and evaluate it in human experiments. The author considers that such a cognitive-architecture-based framework is advantageous in constructing trustful human-agent relations. Academic knowledge implemented in architecture is the result of continuous endeavors in human history. Including the formal knowledge agreed in human society is an essential ingredient of making common ground between humans and artifacts.

\section*{Acknowledgement}
This document summarizes ideas obtained from past collaborative studies with colleagues at Nagoya University, Shizuoka University, collaborators from Panasonic Corp. and Mazda Corp., and members of the Applied Cognitive Modeling Lab (ACML) at Shizuoka university. The author thanks everyone for valuable discussions.

\bibliographystyle{ACM-Reference-Format}
\bibliography{chai}

\end{document}